\newcommand{\be}{\begin{equation}}\newcommand{\ee}{\end{equation}}
\newcommand{\bea}{\begin{eqnarray}}\newcommand{\eea}{\end{eqnarray}}
\newcommand{\nn}{\nonumber}\newcommand{\p}[1]{(\ref{#1})}

\newcommand\s{\scriptscriptstyle}
\newcommand{\E}{\stackrel{\to}{E}}
\newcommand{\M}{\stackrel{\to}{M}}

\documentstyle[12pt]{article}

\begin{document}
\setcounter{page}0
\renewcommand{\thefootnote}{\fnsymbol{footnote}}
\thispagestyle{empty}
\begin{flushright}
JINR E2-97-322 \\
hep-th/9710236 \\
October, 1997
\end{flushright}
\begin{center}
{\large\bf MODIFIED N=2 SUPERSYMMETRY AND \\

\vspace{0.2cm}
FAYET - ILIOPOULOS TERMS}
\vspace{0.5cm} \\
E.A. Ivanov\footnote{E-mail: eivanov@thsun1.jinr.dubna.su}
and B.M. Zupnik\footnote{E-mail: zupnik@thsun1.jinr.dubna.su}
\footnote{On leave of absence from: Institute of Applied Physics,
Tashkent State University, Uzbekistan}
\vspace{0.5cm} \\
{\it Bogoliubov Laboratory of Theoretical Physics, JINR,  \\
141980 Dubna, Moscow Region, Russia} \vspace{1.5cm} \\
{\bf Abstract}
\end{center}
We study peculiarities of realization of $N=2$ supersymmetry in $N=2$
abelian gauge theory with two sorts of $FI$ terms, electric and magnetic
ones, within manifestly supersymmetric formulations via the Mezincescu
and harmonic-analytic prepotentials. We obtain a `magnetic', duality-
transformed superfield form of the $N=2$ Maxwell effective holomorphic
action with standard electric $FI$ term and demonstrate that in such a
system off-shell $N=2$ supersymmetry is inevitably realized in an unusual
Goldstone mode corresponding to the {\it partial} spontaneous breaking
down to $N=1$. On shell, the standard total breaking occurs. In a system
with the two sorts of $FI$ terms, off-shell $N=2$ supersymmetry is
realized in the partial breaking mode both in the electric and magnetic
representations. This regime is retained on shell due to the Antoniadis-
Partouche-Taylor mechanism. We show that the off-shell algebra of $N=2$
supersymmetry in the partial breaking realization is modified on
gauge-variant objects like potentials and prepotentials. The closure of
spinor charges involves some special gauge transformations before any
gauge-fixing.

\newpage
\renewcommand{\thefootnote}{\arabic{footnote}}
\setcounter{footnote}0
\setcounter{equation}0
\section{Introduction}

\hspace{0.5cm}A celebrated mechanism of spontaneous breakdown of rigid
$N=2$ supersymmetry consists in adding a Fayet-Iliopoulos ($FI$) term to
the action of $N=2$ gauge theory. Recently, Antoniadis, Partouche and
Taylor (APT) \cite{APT} have found that the dual formulation of $N=2$
abelian gauge theory (inspired by Seiberg-Witten duality conjecture)
provides a more general framework for such a spontaneous breaking due to
the possibility to define two kinds of the $FI$ terms (see also \cite{Fe}
). One of them (`electric') is standard, while another (`magnetic') is
related to a dual $U(1)$ gauge supermultiplet. APT show that a partial
spontaneous breakdown of $N=2$ supersymmetry to $N=1$ becomes possible,
if one starts with an effective $N=2$ Maxwell action (with some
holomorphic function of $N=2$ superfield strength $W$ as a superfield
Lagrangian) and simultaneously includes two such $FI$ terms.

In this paper we study $N=2$ Maxwell action with the two types of
$FI$-terms and its invariance properties in the framework of
manifestly off-shell supersymmetric $N=2$ superfield formalism, using both
the formulation via the Mezincescu prepotential \cite{Me} and the harmonic
superspace formulation \cite{GIK1}. Our basic observation is that
after duality transformation of a system with even one sort of the $FI$
term, the electric one, off-shell $N=2$ supersymmetry is inevitably
modified, it starts to be realized in a mode with {\it partial}
spontaneous breaking. The dual $N=2$ superfield covariant strength
acquires an unavoidable inhomogeneous term in its supersymmetry
transformation and it can naturally be called $N=2$ {\it Goldstone -
Maxwell superfield} (by analogy with $N=1$ Goldstone - Maxwell superfield
introduced in \cite{BG,Ba} in the nonlinear realizations approach). One of
the dual gaugino is the relevant off-shell Goldstone fermion. On shell
such a system is equivalent to the original system with the standard
`electric' form of the $FI$ term, so after passing on shell the total
breaking of $N=2$ supersymmetry occurs (under some restrictions on the
holomorphic Lagrangian function). The situation is radically changed
after including both types of the $FI$ terms. We show that in this case
off-shell $N=2$ supersymmetry is realized in a partial breaking fashion
in both duality-related formulations, `electric' and `magnetic' ones, with
the electric and magnetic $N=2$ superfield strengths as the relevant
Goldstone-Maxwell superfields. This partial breaking regime is preserved
on shell due to the APT mechanism. We demonstrate how simple the latter
is when using a manifestly $N=2$ supersymmetric formalism. We study the
realization of modified $N=2$ supersymmetry transformations on the
gauge-variant objects ($N=2$ harmonic-analytic prepotential and $N=1$
gauge prepotential) and find that the $N=2$ supersymmetry algebra itself
is also necessarily modified in this case. Namely, the closure of
$N=2$ supercharges contains, besides translations, some special
gauge transformations before any gauge-fixing.

In Sect. 2 we give a brief account of the standard $N=2$ superfield
formulation of abelian $N=2$ gauge theory with the electric $FI$ term,
where off-shell $N=2$ supersymmetry is realized in a customary way.
In Sect. 3 we present a duality-transformed `magnetic' superfield form
of the action of such a theory and demonstrate that $N=2$ supersymmetry
in this representation is necessarily realized off shell in a partial
breaking mode. In Sect. 4 we discuss a general situation with the two
sorts of the $FI$ terms added and show that the regime of off-shell
partial breaking of $N=2$ supersymmetry in this case is stable against
duality transformation and is preserved on shell. In Sect. 5 we briefly
discuss how our observations look in the $N=1$ superfield formulation.
In Sect. 6 we pass to the formulation via the harmonic-analytic $N=2$
prepotential $V^{\s ++}$ and study the modified $N=2$ supersymmetry
transformations and their closure on this fundamental object of $N=2$
gauge theory. We discuss difficulties of constructing minimal couplings
of $V^{\s ++}$ to the matter $q^+$ hypermultiplets in the framework of
such a modified $N=2$ supersymmetry.

\section{$N=2$ gauge theory in ordinary $N=2$ superspace}

\hspace{0.5cm}Superfield constraints of $N=2, D=4$ supersymmetric gauge
 theory were
given for the first time in ref.\cite{GSW}. For the abelian case they
read
\bea
& F^{kl}_{\alpha\beta}=D^k_\alpha A^l_\beta + D^l_\beta A^k_\alpha =
i\varepsilon^{kl}\varepsilon_{\alpha\beta}\overline{W} & \label{A1}\\
& F^{kl}_{\dot{\alpha}\dot{\beta}}=\bar{D}^k_{\dot{\alpha}}
\bar{A}^l_{\dot{\beta}} + \bar{D}^l_{\dot{\beta}} \bar{A}^k_{\dot{\alpha}}
 =i\varepsilon^{kl}\varepsilon_{\dot{\alpha}\dot{\beta}} W & \label{A2}\\
& F^{kl}_{\alpha\dot{\beta}}=D^k_\alpha \bar{A}^l_{\dot{\beta}} +
\bar{D}^l_{\dot{\beta}} A^k_\alpha -i\varepsilon^{kl}
A_{\alpha\dot{\beta}}=0\; &
\label{A3}
\eea
Here $A_{\s M}=(A^i_\alpha,\;\bar{A}_{i\dot{\alpha}},\;
A_{\alpha\dot{\beta}})$ are gauge superfield potentials in the real $N=2$
superspace with the coordinates $z^{\s M}=(x^m, \theta^\alpha_i,
\bar{\theta}^{\dot{\alpha}i})$. They are usually assumed to possess the
$SU(2)$-covariant standard off-shell $N=2$ supersymmetry transformation
 laws
\be
\delta_\epsilon A_{\s M}=i(\epsilon^\alpha_k Q^k_\alpha +
\bar{\epsilon}^{\dot{\alpha}}_k \bar{Q}^k_{\dot{\alpha}}) A_{\s M}\;.
\label{A4}
\ee

The constraints \p{A1} - \p{A3} can be solved either in terms of
unconstrained real prepotential $V^{ik}$ of dimension 2 (the Mezincescu
prepotential \cite{Me}) or in terms of dimensionless analytic
harmonic prepotential $V^{\s++}$ in the framework of the harmonic
superspace approach \cite{GIK1}. We postpone a discussion of the
harmonic-superspace formulation to Sect. \ref{C}, and will firstly
deal with the formulation via $V^{ik}$.

As a consequence of the above constraints and Bianchi identities
the gauge invariant $N=2$ superfield strength $W$ is chiral
\be
\bar{D}_{i\dot{\alpha}}W=0 \label{A6}
\ee
and satisfies the additional constraint
\be
D^{ik} W - \bar{D}^{ik} \overline{W} = 0~, \label{A7}
\ee
where the standard notation for bilinear combinations of the
spinor derivatives $D^i_\alpha$ and $\bar{D}_{i\dot{\alpha}}$
is used, $D^{ik}=D^{i\alpha} D^k_{\alpha}  $ and $\bar{D}^{ik}=
\bar{D}^i_{\dot{\alpha}} \bar{D}^{k\dot\alpha} $. The constraint
\p{A7} is the reality condition implying the auxiliary component
of $N=2$ Maxwell multiplet,
\be \label{defX}
X^{ik} \equiv -{1\over 4}D^{ik}W|_0\;,
\ee
to be real
\be \label{realX}
(X^{ik})^\dagger = \varepsilon_{il}\varepsilon_{km}X^{lm} = X_{ik}
\ee
(the symbol $|_0$ means restriction to the lowest, $\theta$, $\bar\theta$
-independent component of $N=2$ superfield).

Both these constraints on $W$ can be solved through the Mezincescu
prepotential \cite{Me}
\be
W_{\s V} = (\bar{D})^4 D_{ik} V^{ik}~.
\label{A8}
\ee

It should be emphasized that $N=2$ gauge theory  can be fully specified
by the covariant strength superfield $W$ subjected to the constraints
\p{A6}, \p{A7} (or a generalization of the latter, see next Sections).
So we can deal entirely with $W$ and $V^{ik}$ as the basic objects of
the theory and not care about their geometric origin.

A holomorphic effective action for the abelian (electric) prepotential
$V^{ik}$ has the following form
\be
S(V)=\frac{i}{4}\int d^4 x d^4 \theta {\cal F}(W_{\s V}) +
\mbox{c.c.}\;.
\label{A9}
\ee
Here, ${\cal F}(W_{\s V}) $ is some holomorphic function and
$d^4 \theta = (D)^4$. The prepotential $V^{ik}$ can be also used to
construct a gauge-invariant $FI$ term which breaks the
$SU(2)$-automorphism symmetry and is capable to induce a spontaneous
breakdown of $N=2$ supersymmetry
\be
S_{\s FI}(V)=\int d^{12}z E_{ik}V^{ik}\;, \quad S(V) \rightarrow
S_{\s E}(V) = S(V) + S_{\s FI}(V)~.  \label{A10}
\ee
Here, $E^{ik}=i\E(\stackrel{\to}{\sigma})^{ik}$ is a
$SU(2)$ triplet of constants satisfying the same reality condition
\p{realX} as the auxiliary field $X^{ik}$:
\be
(E^{ik})^\dagger \equiv E^\dagger_{ik}=
\varepsilon_{il}\varepsilon_{kn}E^{ln}=E_{ik}~, \quad
\mbox{or} \quad \E^\dagger = \E\;.
\label{reality}
\ee
Note that for any real vector $\E \neq 0$ the matrix $E^{ik}$
is non-degenerate
\be
\mbox{Det}E^{ik}\sim \E^2 \neq 0~. \label{nondeg}
\ee

The superfield equation of motion following from the action
$S_{\s E}(V)$ by varying $V^{ik}$ reads
\be
D^{kl}{\cal F}_{\s W}(W_{\s V}) - \mbox{c.c.} = \left[ \tau(W_{\s V})
D^{kl}W_{\s V}
+
\tau^\prime (W_{\s V}) D^{k \alpha} W_{\s V}D^l_{\alpha} W_{\s V}
\right] - \mbox{c.c.}=4iE^{kl}~, \label{eqmFI1}
\ee
where ${\cal F}_{\s W}=\partial{\cal F}/\partial W$ and the standard
notation for the effective coupling constant and its derivative is used
\be
\tau(W)=\frac{\partial^2{\cal F}}{\partial W^2}=\tau_1 + i\tau_2 \;\;
(\tau_2 > 0)~,
\quad
\tau^\prime(W)=\frac{\partial^3{\cal F}}{\partial W^3}~.
 \label{effcon}
\ee
Hereafter, it is assumed that the $SU(2)$ indices in the c.c. pieces are
put in a proper position with the help of skew-symmetric tensors, e.g.
$$\overline{X^{ik}}=\varepsilon^{ij} \varepsilon^{kl}X^\dagger_{jl}~.$$

A possibility of spontaneous breakdown of $N=2$ supersymmetry
by the $FI$ term is related to the possibility to
have a non-zero vacuum solution for the auxiliary component $X^{ik}$
in this case
\be
<X^{ik}>\equiv x^{ik} \sim\;E^{ik}~. \label{realaux}
\ee
Provided that such a solution exists and corresponds to a stable
classical vacuum, there appears an inhomogeneous term in the
on-shell supersymmetric transformation law of the $N=2$ gaugino
doublet $\lambda^{i\alpha}$
\be
\delta \lambda^{i\alpha} \sim \epsilon^\alpha_k E^{ik}\;,
\ee
$\epsilon^{\alpha}_k$ being the transformation parameter.
Thus there are Goldstone fermions in the theory, which is a standard
signal of spontaneous breaking of $N=2$ supersymmetry.

It is easy to see that for any non-degenerate matrix $E^{ik}$
{\it both} $\lambda^{1\alpha}, \lambda^{2\alpha}$ are shifted by
independent parameters, and so they both are Goldstone fermions in
this case. Thus, with the standard $FI$ term, only {\it total}
spontaneous breaking of $N=2$ supersymmetry can occur.
Recall that the inhomogeneous pieces in the transformation laws of
$\lambda^{\alpha i}$ appear as a result of solving the equation
of motion for $X^{ik}$, so it is natural to assign the term
`{\it on-shell} Goldstone fermions' to these fermionic fields.

In order to get a feeling in which cases the $FI$ term indeed
generates a spontaneous breaking of $N=2$ supersymmetry, let us examine
whether a non-trivial vacuum background solution with constant values of
the auxiliary component $<X^{ik}>= x^{ik}$ and the scalar
field $<\Phi>\equiv <W_{\s V}>|_{0}= a$ exists. So, we
choose the ansatz
\be  \label{anz1}
<W_{\s V}>_{\s 0} = a + (\theta_i\theta_k)\; x^{ik}\;,
\ee
where $(\theta_i\theta_k) = \varepsilon_{\alpha\beta}
\theta^\alpha_i\theta^\beta_k$, and substitute it into the equation of
motion \p{eqmFI1}. Using the identity $D^{ik}(\theta_j\theta_l)=
-2(\delta_j^i \delta_l^k + \delta_j^k\delta_l^i)$ we get two
independent equations
\bea
& x^{ik}\;\tau_2(a) = - {1\over 2} E^{ik}~,& \label{scal1} \\
& \tau^\prime(a)\; x^{ik}x_{ik} = \tau^\prime(a)\; |x|^2= 0 &
\label{scal2}
\eea
(the second one most directly follows from the equation
$(D)^4 {\cal F}_{\s W} \sim \Box \overline{\cal F}_{\s W}$
which can be obtained by applying $D_{kl}$ to eq. \p{eqmFI1}).

A constant solution $x^{ik} \sim E^{ik}$ to eqs. \p{scal1}, \p{scal2}
evidently exists only if $\tau^\prime = 0$, that corresponds to the
quadratic Lagrange function
${\cal F}(W_{\s V}) \sim W_{\s V}^2$, i.e. to the free $N=2$
Maxwell theory.
Thus for non-trivial functions ${\cal F}$ in the action
$S (W) + S_{\s FI}$, the coupled set of equations of motion
for physical and auxiliary bosonic fields admits no constant regular
solutions which could trigger a spontaneous breaking of $N=2$
supersymmetry. This fact was firstly noticed in ref. \cite{GO}. In the
same reference, it was also shown that a stable vacuum with a constant
nonvanishing $X^{ik}$ and, hence, spontaneously broken $N=2$
supersymmetry exists in a system of at least two $N=2$, $U(1)$ gauge
superfields with the $FI$ term for one of them. As is discussed in the
next Section, a spontaneous breakdown with a non-trivial function
${\cal F}$ and yet one gauge superfield becomes possible when choosing
a more general $N=2$ Maxwell action with two different sorts of $FI$
terms, `electric' and `magnetic' \cite{APT,Fe}. Moreover, in this case
a {\it partial} breaking of $N=2$ supersymmetry down to $N=1$ can occur.

\setcounter{equation}{0}
\section{Dual form of $FI$ term and modification of
$N=2$ supersymmetry \label{B}}

\hspace{0.5cm}Now we turn to discussing the spontaneous breakdown of
$N=2$ supersymmetry within dual formulations of the $N=2$ Maxwell
effective action. In constructing such formulations we follow
the lines of refs. \cite{APT,SW,He}.

The passing to the dual description goes through some
intermediate `master' action with an enlarged set of superfields.
It involves a chiral and otherwise unconstrained `electric' superfield
strength $W$ and some constrained `magnetic' superfield strength .
Both the original and dual formulations follow from this `master' action
upon varying it with respect to proper superfields.

To get the `master' action, let us add the constraint
(\ref{A7}) to the action \p{A9} with the help of an unconstrained
$N=2$ superfield Lagrange multiplier $L_{ik}$
\bea
S(V) \rightarrow S(W,L)= S(W) + \frac{i}{4}
\int d^{12}z L_{ik}(\bar{D}^{ik}\overline{W} -
D^{ik}W) \equiv S(W) + S_{\s L}\;,
\label{B2}
\eea
where $S(W)$ is obtained via the substitution
$W_{\s V} \rightarrow W$ in \p{A9}. Thus, the action $S(W,L)$
includes an unconstrained real superfield $L_{ik}$ and
a chiral superfield $W$ that is otherwise arbitrary.

Varying $L^{ik}$ yields the constraint \p{A7} and hence leads us back
to the `electric' action \p{A9} written in terms of $W_{\s V}$, eq.
 \p{A8}. On the other hand, one can rewrite \p{B2} as an
integral over the chiral subspace \cite{APT,He}
\bea
S(W,L)&=& \frac{i}{4}\int d^4x d^4\theta
[{\cal F}(W)- W W_{\s L}]
 + \mbox{c.c.}~,
\label{B3} \\
W_{\s L} &=& (\bar{D})^4 D_{ik} L^{ik} \;. \label{B4}
\eea
The newly introduced chiral object $W_{\s L}$ by construction satisfies
the same constraint \p{A7} as $W_{\s V}$, i.e.
\be
D^{ik}W_{\s L} - \bar D^{ik}\overline{W}_{\s L}= 0\;, \label{Dconstr}
\ee
and is expressed via $L^{ik}$
just in the same fashion as $W_{\s V}$ via the Mezincescu
prepotential $V^{ik}$. Therefore it is natural to think of $W_{\s L}$
as the dual or `magnetic' $N=2$ Maxwell superfield strength, and the
Lagrange multiplier $L^{ik}$ as the dual or `magnetic' prepotential.

In order to obtain a  `magnetic' representation of the $N=2$ Maxwell
action, one should eliminate $W$ from the `master' action \p{B2} by
varying the latter with respect to this superfield. As a result one gets
an algebraic equation
\be
 {\cal F}_{\s W}=W_{\s L} \label{B5}
\ee
that allows one to express $W$ in terms of $W_{\s L}$
\bea
&W = W(W_{\s L})\;,
& \label{B6} \\
&\partial W / \partial W_{\s L} =[\partial W_{\s L}/\partial W ]^{-1}=
(\tau(W))^{-1} \equiv - \hat{\tau}(W_{\s L})~.&
\label{B6b}
\eea
After this one arrives at the magnetic representation of the $N=2$
Maxwell action
\be
S(L)= \frac{i}{4}\int d^4x d^4\theta \hat{\cal F}(W_{\s L})
+ \mbox{c.c.} \;,
\label{B7}
\ee
with the new dual holomorphic Lagrangian function
\be  \label{hatF}
\hat{\cal F}(W_{\s L}) \equiv {\cal F}[W(W_{\s L})]
- W_{\s L}\;W(W_{\s L})\;.
\ee
The `magnetic' equation of motion has the following simple form:
\be
D^{ik} \hat{\cal F}{}' - \mbox{c.c.} =
\left( \hat{\tau} D^{ik} W_{\s L}
+ \hat{\tau}^\prime D^{k\alpha} W_{\s L}D^l_{\alpha} W_{\s L}\right) -
\mbox{c.c.} = 0~.
  \label{magneq}
\ee

Thus, the functional $S(W,L)$ \p{B2} defines the duality transformation
between the `electric' and `magnetic' forms of the $N=2$ gauge theory
action
\be
S(V) \leftrightarrow S(W,L) \leftrightarrow S(L)~.
\ee

How to get the dual form of the $FI$-term (\ref{A10})?
Recall that in the original `electric' representation
it is constructed using the prepotential $V_{ik}$, the object which
appears as the solution to the constraint (\ref{A7}) and which is
certainly lacking in the formalism with the chiral and otherwise
unconstrained superfield $W$ and the dual superfield strength $W_{\s L}$.

To answer this question, let us come back to the `master' action \p{B2}
and extend it by the term
\be
S_e =
-\frac{1}{8}\int d^{4}x d^4\theta E^{ik} (\theta_i\theta_k)W +
\mbox{c.c.}\;, \quad S(W, L) \rightarrow S_{\s E} (W,L) = S(W,L) + S_e\;.
\label{B8}
\ee
Note that the constants $E^{ik}$ in $S_e$, without loss of generality,
can be chosen real; their possible imaginary parts
can always be absorbed into a redefinition of $W_{\s L}$ or $L^{ik}$
without affecting the reality properties of these superfields.
The term $S_e$ becomes just (\ref{A10}) after the substitution
$W \rightarrow W_{\s V}$, i.e. after passing to the
`electric' representation, and hence it can be regarded as a `disguised'
form of the standard electric $FI$ term. Its dual `magnetic' form can
now be obtained by passing to the `magnetic' representation of the
extended action $S_{\s E}(W,L)$ by eliminating $W$ from it, like this
was done for the action $S(W,L)$.

However, at this step one encounters a trouble. We observe that
$S_e$ is not invariant under the standard $N=2$
supersymmetry transformations unless $W$ is subjected to the
constraint \p{A7}. The invariance of the full action can be
restored (before imposing \p{A7}, i.e., varying with respect to
$L^{ik}$) by means of the following redefinition of the off-shell
transformation law of the dual superfield strength
\be
\delta_\epsilon W_{\s L} = i(\epsilon_k\theta_l)E^{kl} +
 i(\epsilon Q + \bar{\epsilon}\bar{Q})W_{\s L}\;,
 \label{B9}
\ee
where $Q^i_\alpha, \bar Q^i_{\dot\alpha}$ are standard $N=2$
supersymmetry generators. Note that the appearance of
the $SU(2)$-breaking shift in eq.(\ref{B9}) is still compatible
with the constraint \p{Dconstr} for $W_{\s L}$, thanks to the
relation $D^{ij}(\epsilon_k\theta_l)=0$.

This modified transformation law still has the space-time
translations as the off-shell closure, but implies a Goldstone-type
transformation for the fermionic component $D^{\alpha i}W_{\s L}
\equiv \lambda^{\alpha i}_{\s L}$ (i.e.,
the `magnetic' photino)
\be
\delta \lambda^{\alpha i}_{\s L} \sim i \epsilon^\alpha_k E^{ik} \;.
\ee
This inhomogeneous transformation is valid off shell, before
using the equations of motion, therefore $\lambda_{\s L}^{\alpha i}$
can be called {\it off-shell} Goldstone fermions.

With the definition \p{B9},
inhomogeneous pieces are present in {\it both} supersymmetry
transformations, so at first sight we are facing the phenomenon of total
off-shell spontaneous breaking of $N=2$ supersymmetry in this case. It is
not so, however. Namely, let us show that by a proper shift of the real
auxiliary field of $W_{\s L}$
\be
W_{\s L} \rightarrow \widetilde{W}_{\s L} = W_{\s L} +
{1\over 2}(\theta_i\theta_k)C^{ik},         \label{shiftWL}
\ee
one can restore a homogeneous transformation law with respect to
{\it one} of two $N=1$ supersymmetries present in $N=2$
supersymmetry (it is easy to find the appropriate redefinition of
$L^{ik}$). The newly defined object $\widetilde{W}_{\s L}$ transforms
as follows
\be
\delta_\epsilon \widetilde{W}_{\s L} = (\epsilon_k\theta_l)(C^{kl}
+iE^{kl})+ i(\epsilon Q + \bar{\epsilon}\bar{Q})\widetilde{W}_{\s L}\;.
\label{B9til}
\ee
One can always choose $C^{ik}$ so that
\be  \label{degener1}
\mbox{det}\;(C+iE) = 0~.
\ee
Indeed, this condition amounts to requiring $C^{ik}$ to be orthogonal
to $E^{ik}$ and to have the same norm
\be \label{conds}
(a)\;E^{ik}C_{ik} = 0, \quad (b)\;|E| = |C|~.
\ee
It is easy to find a general solution to these equations. E.g.,
for the two different choices of the $SU(2)$ frame:
\be
(i)\;E^{\s 12} \neq 0,\; E^{\s 11} = E^{\s 22} = 0~;\quad
(ii)\;E^{\s 12} = 0,\; E^{\s 11} = E^{\s 22}  \label{frame1}
\ee
we have
\be
(i)\;C^{\s 12} = 0, \; C^{\s 11}C^{\s 22} = |E^{\s 12}|^2\;;\quad
 (ii)\; C^{\s 12} = 0,\; C^{\s 11} = \pm iE^{\s 11},\;C^{\s 22} =
\mp iE^{\s 11}~.
\label{frame2}
\ee
Note that in the second case we have fixed the residual $U(1)$
freedom up to a reflection.

Eq.\p{degener1} means that $C^{ik} + iE^{ik}$ is a degenerate
symmetric $2\times 2$ matrix, so it can be brought to the form with only
one non-zero entry (ii). As a result, $\widetilde{W}_{\s L}$ is actually
shifted under the action of only {\it one} linear combination of the
modified $N=2$ supersymmetry generators $\hat{Q}^{1,2}_\alpha$, while
under the orthogonal combination it is transformed homogeneously.
The same is true of course for the physical fermionic components:
only {\it one} their combination is the genuine off-shell Goldstone
fermion. All these options are related by some $SU(2)$
transformations (they can be continuous or discrete). For instance,
in the case (ii) in eq. \p{frame2} the $\epsilon_2^\alpha$ or
$\epsilon_1^\alpha$  supersymmetries are broken for the first or
second choices of the sign, respectively.

Thus we arrive at the important conclusion: in the dual, `magnetic'
representation of $N=2$ Maxwell theory with $FI$ term $N=2$
supersymmetry is realized {\it off shell} in a mode with {\it
partial spontaneous breaking}, so that some $N=1$ supersymmetry
remains unbroken.

It should be specially pointed out that, contrary to the original,
electric representation of the $FI$ term, in the dual representation
the phenomenon of spontaneous breaking of $N=2$ supersymmetry occurs
already {\it off shell}, irrespective of the form of the holomorphic
Lagrangian function $\hat{\cal F}(W_{\s L})$. This situation resembles
the nonlinear realization approach \cite{CWZ,BG2,BG,Ba} where one
introduces from the very beginning, according to some prescription,
inhomogeneously transforming Goldstone fields or superfields which
express in a most pure way an effect of off-shell spontaneous breaking
of some symmetry or supersymmetry.

Partial spontaneous breaking of $N=2$ supersymmetry in the framework of
the nonlinear realizations approach, with the relevant Goldstone fermion
placed into chiral or vector $N=1$ multiplets, was considered in
ref. \cite{BG2,BG,Ba}.
Using the terminology of \cite{BG,Ba} and taking into account that
$W_{\s L}$ contains Goldstone fermion components, it is natural to call
this inhomogeneously transforming superfield the $N=2$
{\it Goldstone-Maxwell} ($GM$) superfield.
In what follows we will deal with this superfield, keeping in mind that
one can always pass to $\widetilde{W}_{\s L}$, eq. \p{B9til}, in terms of
which the phenomenon of the {\it off-shell partial $N=2$ supersymmetry
breaking} becomes manifest.

One can go a step further and wonder how the modified transformations
are realized on the gauge-variant objects: gauge potentials and
prepotentials. In Sect. \ref{D} we will present the modified
supersymmetry transformation of the $N=1$ $GM$ prepotential $V$.
In Sect. \ref{C}, we will also give how the modified $N=2$ supersymmetry
is realized on the harmonic prepotential and study a
modified supersymmetry algebra in this realization. As we will see,
the $GM$ mechanism of spontaneous breaking implies an essential
modification of the algebra of $N=2$ supersymmetry transformations on
gauge superfields. Note that the transformation properties of the
magnetic gauge connections $A_{\s M}$ are also appropriately modified by
inhomogeneous terms containing $E^{ik}$, so as to preserve the covariance
of the magnetic analogs of the constraints \p{A4}.

As was already mentioned, after passing to the `electric' representation
of $S_{\s E}(W,L)$ by varying it with respect to $L^{ik}$ one gets
the action \p{A9} accompanied by the standard $FI$ term \p{A10},
both being written in
terms of the `electric' prepotential $V^{ik}$. This prepotential and its
covariant chiral strength $W_{\s V}$ possess standard $N=2$ supersymmetry
transformation properties, so the modification of $N=2$ supersymmetry in
the action $S_{\s E}(W,L)$ is to some extent an artifact related to
the insertion of the constraint \p{A7} into the action and the appearance
of additional superfield $L^{ik}$ there. However, this modified $N=2$
supersymmetry is retained after putting $S_{\s E}(W,L)$ into the pure
`magnetic' representation by eliminating the superfield $W$. Thus it is a
characteristic unavoidable feature of the dual (`magnetic') off-shell
formulation of $N=2$ Maxwell action with the $FI$ term.

To get the precise form of such a dual action, let us combine the $L^{ik}$
(or $W_{\s L}$) piece $S_{\s L}$ in the action $S_{\s E}(W,L)$, eq.
\p{B8}, together with $S_e$ into the following mixed term
\be
\hat{S}_{\s L}=S_{\s L} + S_e= -{i\over 4}\int d^4x d^4\theta  W
\hat{W}_{\s L}+ \mbox{c.c.}\;,
\label{B10}
\ee
where a shifted $GM$-superfield $\hat{W}_{\s L}$
with a homogeneous $N=2$ supersymmetry transformation law
was introduced
\bea
\hat{W}_{\s L} &=& -(i/2)(\theta_k\theta_l)E^{kl}+W_{\s L}~,
 \label{B11} \\
\delta_\epsilon \hat{W}_{\s L} &=& i(\epsilon Q + \bar{\epsilon}
\bar{Q})\hat{W}_{\s L}\;. \label{B11b}
\eea
The net difference between $\hat{W}_{\s L}$ and $W_{\s L}$
is that the auxiliary scalar field component of the former,
$$
\hat{Y}^{ik} \equiv -{1\over 4}D^{ik}\hat{W}_{\s L}|_0~,
$$
contains a constant imaginary part proportional to $E^{ik}$, while
the auxiliary component ${Y}^{ik}$ of $W_{\s L}$
is real by construction (eq. \p{B4}). One has
\be
 \hat{Y}^{kl} = Y^{kl} - (i/2)E^{kl}~.
\label{auxmagn}
\ee
By its definition, the quantity $\hat{W}_{\s L}$
obeys the modified constraint
\be
D^{kl}\hat{W}_{\s L} - \mbox{c.c.} = 4iE^{kl}~,
\label{B12}
\ee
demonstrating the presence of the imaginary
constant part just mentioned.

Clearly, passing to $\hat{W}_{\s L}$ does not redefine the `magnetic'
gaugini and does not affect the interpretation of their one combination
as a Goldstone fermion with an inhomogeneous piece in the off-shell
transformation law.

It is easy to get a purely `magnetic' form of the modified `master'
action $S_E(W,L)$, eq. \p{B8}. Varying $W$ now yields the modified
equation
\be
{\partial {\cal F}\over \partial W} =
\hat W_{\s L}\;, \label{mod}
\ee
which leads to the following $\theta$-dependent modification of the
magnetic action \p{B7}
\be
\tilde{S}(L) =
\frac{i}{4}\int d^4x d^4\theta \hat{\cal F}(\hat{W}_{\s L}) +
\mbox{c.c.} \;,\label{newmagn}
\ee
with $\hat{\cal F}$ defined by eq. \p{hatF}.

Thus in the dual `magnetic' formulation the whole effect of the
original `electric' $FI$ term \p{A10} is the appearance of
$\theta$-dependent terms in the action, with coefficients proportional
to the $SU(2)$ breaking constants $E^{ik}$. Despite this explicit
$\theta$ dependence, the action is still $N=2$ super-invariant due to the
modification \p{B9} of the transformation law of $W_{\s L}$.

The appropriate equation of motion follows from \p{magneq} by
substituting there $W_{\s L} \rightarrow \hat{W}_{\s L}$ and taking
account of the modified constraint \p{B12}.
Analyzing this equation, it is straightforward to show that the dual
action \p{newmagn} leads to the same vacuum structure as the original
`electric' action with the $FI$ term \p{A10}.
In particular, a Poincar\'e-invariant vacuum with spontaneously broken
$N=2$ supersymmetry exists only for
$\hat{\tau}{}' = 0$, i.e. for a free theory, like in the electric
representation (this restriction is equaivalent to $\tau{}' = 0$,
of course).

Thus on shell in the `magnetic' representation
we again face the total spontaneous breaking of $N=2$ supersymmetry,
despite the fact that off-shell $N=2$ supersymmetry is realized in
the mode with partial spontaneous breaking.

To summarize, in the `magnetic' representation of
$N=2$ Maxwell theory with a single $FI$ term there occurs a
partial spontaneous breaking of $N=2$ supersymmetry off shell which
goes over into the total one after passing on shell. Thus, a crucial
difference from the original, `electric' representation of
the same theory is that in the `magnetic' case the effect of spontaneous
supersymmetry breaking has an unavoidable `inborn' off-shell constituent
corresponding to the partial breaking mode. This new phenomenon is related
to the modification of off-shell realization of $N=2$ supersymmetry after
performing a duality transform of the `electric' $FI$ term. Passing on
shell changes the type of spontaneous breakdown, promoting it to the
total one (simultaneously imposing a severe constraint on the admissible
form of the holomorphic Lagrangian).

In the next Section we will see that adding of some different kind of
the $FI$ term radically changes this interplay between the off-shell and
on-shell constituents of spontaneous $N=2$ supersymetry breaking. Namely,
there appears a stable vacuum ensuring the phenomenon of partial
supersymmetry breaking to retain on shell, too.

\setcounter{equation}{0}

\section{$N=2$ supersymmetry in the presence
of electric and magnetic $FI$ terms \label{BB}}

\hspace{0.5cm}New possibilities come out if we simultaneously include
two alternative mechanisms of the spontaneous supersymmetry
breaking, the previously discussed one and a new one discovered
in \cite{APT}. It consists in extending the intermediate action
$S_{\s E}(W,L)$ \p{B8} by a superfield `magnetic' $FI$-term
\be
S_{m}=\frac{1}{8}\int d^{4}x d^4\theta M^{kl} (\theta_k\theta_l)W_{\s L}
+ \mbox{c.c.} =-\int d^{\s 12}z M^{kl}L_{kl}~,
\label{B13}
\ee
with $M^{ik}$ being another triplet of real constants. The $W_{\s L}$-form
of $S_{m}$ is invariant under the Goldstone-type transformation (\ref{B9})
since $\int d^4 \theta (\theta)^3 = 0$ (a contribution from the shift of
explicit $\theta$s in \p{B13} disappears by rewriting $W_{\s L}$ through
$L^{ik}$, restoring the full integration measure and integrating by parts
). One can show that the inhomogeneous piece in the transformation of
$L_{ik}$ does not contain terms higher than those of 7th order in the
spinor coordinates, so the invariance of the second representaton of
$S_m$ in \p{B13} is also guaranteed,
$\int d^8\theta \delta_\epsilon L_{ik}=0$.

As in the previous consideration, one can pass from this modified
intermediate action to its either `electric' or `magnetic'
representations, varying it with respect to $L^{ik}$ or $W$,
respectively.

When one descends to the `electric' representation (by varying $L^{ik}$),
the only effect of magnetic $FI$ term is the modification of the
constraint \p{A7} on $W$. The modified constraint reads
\be
D^{ik}W - \bar D^{ik}\overline{W} = 4iM^{ik}~  .
\label{B14}
\ee
It suggests the redefinition
\be \label{B14b}
W = W_{\s V} - {i\over 2}(\theta_i\theta_k)M^{ik}~,
\ee
with $W_{\s V}$ satisfying eq. \p{A7} and hence given
by eq. \p{A8} \footnote{
Note that
the $SU(2)$-noninvariant constraint \p{B14} with an arbitrary
$M^{ik}$ (equally as its `magnetic' analog \p{B12}) is consistent with the
constraint of ref. \cite{APT} which can be written in the following form:
$$
D^{ik}  D_{ik} W=D^{ik} \bar{D}_{ik} \overline{W}\sim \Box \overline{W}~.
$$
With using this constraint (or its magnetic analog), the
constants $M^{ik}$ or $E^{ik}$ appear as integration constants, while in
the approach we keep to both these sets are present in the action from
the beginning as some moduli of the theory. The constant terms in the
superfield constraints have been considered also in a model with the
partial breaking of $D=1$ supersymmetry \cite{IKP}.}
, and means the appearance
of the constant imaginary part $-(i/2)M^{ik}$ in the auxiliary field of
$W$,
$$
\hat{X}^{ik} \equiv -{1\over 4} D^{ik} W|_0 = X^{ik} - {i\over 2}M^{ik}~,
\quad X^{ik} \equiv -{1\over 4} D^{ik} {W_{\s V}}|_0~.
$$

The inclusion of magnetic $FI$ term cannot change the transformation
properties of $W$ under $N=2$ supersymmetry which are standard. Then the
relation \p{B14b} requires modifying the transformation
law of $W_{\s V}$, and, respectively, $V^{ik}$ on the pattern of eq.
\p{B9}
\be
\delta_\epsilon W_{\s V} = i(\epsilon_k\theta_l)M^{kl} +
 i(\epsilon Q + \bar{\epsilon}\bar{Q})W_{\s V}
\label{modtraV}
\ee
(we do not give how the transformation law of $V^{ik}$ is changed, it
is easy to find this modification). In other words, $N=2$ supersymmetry
is now realized in a Goldstone-type fashion in the `electric'
representation as well, but with $M^{ik}$ instead of $E^{ik}$ as the
`structure' constants. Thus we see that the adding of magnetic $FI$
term modifies $N=2$ supersymmetry in the electric representation
quite similarly to what happens in the magnetic representation after
adding the standard `electric' $FI$ term. So, when
both $FI$ terms are present, $N=2$ supersymmetry is realized in the
Goldstone mode in both representations, `electric' and `magnetic',
and there is no way to restore the standard $N=2$ supersymmetry off
shell. In particular, $N=2$ transformations of the standard `electric'
gauge connections introduced by eqs. \p{A1} -\p{A4} acquire inhomogeneous
$SU(2)$ breaking terms proportional to $M^{ik}$ (note that in the r.h.s.
of the constraints \p{A1} -\p{A2}, by definition, just the covariant
strength $W_{\s V}$ appears). The same arguments as in the previous
Section show that $N=2$ supersymmetry in both representations is realized
off shell in {\it the partial breaking mode}.

An effect of the magnetic $FI$ term in the magnetic representation
is a further modification of the equation of motion for $W_{\s L}$:
the term $4iM^{ik}$ appears in its r.h.s. Of course,
this is related to the fact that the reality constraint for $W_{\s V}$ is
the equation of motion for the dual superfield strength, and vice versa.

The issue of on-shell spontaneous breaking of $N=2$ supersymmetry
in the presence of electric and magnetic $FI$ terms can be analyzed
both in the electric and magnetic versions of the full intermediate
action
\be
S_{\s (E,M)} = S_{\s E}(W,L) + S_m~,
\ee
with the same final conclusions. In order to be closer to the original
paper \cite{APT}, we prefer to do this in the electric representation,
with the prepotential $V^{ik}$ and its covariant strength $W_{\s V}$ as
the basic entities.

A general electric effective action of the abelian gauge model with the
$(E,M)$- mechanism of the spontaneous breaking can be obtained by
substituting the expression for $W$, eq.\p{B14b}, into the action \p{A9}
and supplying the latter with the electric $FI$ term \p{A10}:
\be
S_{\s (E,M)}=\left [\frac{i}{4}\int d^4x d^4\theta {\cal F}(W)
 + \mbox{c.c.} \right ] + \int d^{12}z E_{ik} V^{ik}~, \quad
 W = W_{\s V} - (i/2)(\theta_i\theta_k)M^{ik}~.
\label{B14c}
\ee

The superfield equation of motion in the electric
representation of the $(E,M)$-model reads
\be
[\tau D^{kl} W  + \tau^\prime D^{k\alpha} W
D^{l}_\alpha W ] - \mbox{c.c.} = 4iE^{kl}~. \label{EMeq}
\ee
The corresponding equation for the auxiliary component
is as follows:
\be
2\tau_2(a)X^{kl} = \tau_1(a) M^{kl} - E^{kl}~. \label{EMcomp}
\ee
Then we take for $<W_{\s V}>$ the ansatz \p{anz1} and substitute it into
\p{EMeq} with taking account of the relation \p{B14b}. Besides the
expression for the vacuum value of auxiliary field $x^{kl} = <X^{kl}>$
\be \label{Ysolut}
x^{kl} = \frac{1}{2\tau_2(a)}\left( \tau_1(a) M^{kl} - E^{kl} \right)~,
\ee
one gets, from vanishing of the coefficient before $(\theta_i\theta_k)$,
the following generalization of eq. \p{scal2}
\be \label{IIequ}
\tau^\prime\;\hat{x}^{ik} \hat{x}_{ik} = 0~, \quad
\hat{x}^{ik} = <\hat{X}^{ik}> = x^{ik} - (i/2)M^{ik}~.
\ee
A crucial new point compared to \p{scal2} is that the vector
$\hat{x}^{ik} = x^{ik} - (i/2)M^{ik}$ is {\it complex}, so the vanishing
of its square does not imply it to vanish. As a result, besides the
trivial solution $\tau^\prime = 0$, eq. \p{IIequ} possesses the nontrivial
one
\be \label{degener}
\tau^\prime \neq 0~, \quad \hat{x}^{ik}\hat{x}_{ik} = 0 \quad (\mbox{or}
\;\;\;\mbox{det}\; \hat{x} = 0)~.
\ee
This solution, after substitution of the expression \p{Ysolut}, amounts
to the following relations between $\tau_{1,2}(a)$ and the $SU(2)$
breaking parameters $E^{ik}, M^{ik}$
\be
\tau_1(a)=\frac{\E\M}{\M^2}~,\quad |\tau_2(a)|=
\frac{\sqrt{\E^2\M^2-(\E\M)^2}}{\M^2} = \frac{|\E \times \M|}{\M^2}~.
\label{taurestr}
\ee
This is just the vacuum solution that was found in \cite{APT} by
minimazing the scalar field potential in the component version of the
action \p{B14c}. It triggers a {\it partial} spontaneous breaking of
$N=2$ supersymmetry down to $N=1$.

Let us show in two equivalent ways that at this point of moduli space
the half of supersymmetry indeed continues to be unbroken on shell.

It follows already from the nilpotency of the complex matrix
$\hat{x}^{ik}$ (eq. \p{degener}) that by a proper rotation it can be
brought into the form with only one nonvanishing element. As a result,
only one linear combination of the inhomogeneously transforming Goldstone
fermions
$$
\delta \lambda^{\alpha i} = \epsilon^\alpha_k \hat{x}^{ik}
$$
retains an inhomogeneous piece in its transformation.

To be more explicit, let us choose the $SU(2)$ frame so as to leave
in $E^{ik}$, $M^{ik}$ only three independent components, e.g.
\bea  \label{frame}
M^{\s12} &=& E^{\s12} = 0, \quad M^{\s11} = M^{\s22} \equiv m, \quad
\mbox{Re}\;E^{\s11} \equiv -e, \quad
\mbox{Im}\;E^{\s11} \equiv \xi \Rightarrow \nn \\
\tau_1 &=& - \frac{e}{m}, \quad |\tau_2| = \frac{|\xi|}{|m|}~.
\eea
In this frame
\be
\hat{x}^{\s12} = 0, \quad \hat{x}^{\s11} = {i\over 2\tau_2} \left(\xi -
m \frac{|\xi|}{|m|}\right), \quad  \hat{x}^{\s22} =
 -{i\over 2\tau_2}\left(\xi + m \frac{|\xi|}{|m|}\right)~.
\ee
One observes that either $\hat{x}^{\s11}$ or $\hat{x}^{\s22}$ is zero,
depending on the relative sign of moduli $m$, $\xi$. Choosing, for
definiteness, $m>0, \xi >0$, one finds
\be \label{entries}
\hat{x}^{\s12} = \hat{x}^{\s11} = 0, \quad \hat{x}^{\s22} = -i m~.
\ee
As a result, only $\lambda^{\alpha 2}$ contains an inhomogeneous term
$\sim m\;\epsilon^\alpha_2 $ in its transformation, and it is the only
genuine on-shell goldstino. So, the $\epsilon^\alpha_2$ supersymmetry is
broken, while the $\epsilon^\alpha_1$ one is not.

An equivalent way to reach the same conclusions is to study the action
of the modified $N=2$ supersymmetry generators $\hat{Q}^i_\alpha$
on the vacuum superfield
\be
<W_{\s V}> = a + (\theta_i\theta_k)x^{ik}~.
\ee
In accord with eqs. \p{modtraV}, \p{B14b} one gets
\be \label{genvac}
\hat{Q}^i_\alpha <W_{\s V}> =
Q^i_\alpha <W_{\s V}> + \theta_{\alpha l}M^{il} = Q^i_\alpha <W>~,
\ee
where
\be
<W> = a + (\theta_i\theta_k)\hat{x}^{ik}~. \label{vacW}
\ee
Substituting the entries \p{entries} (we are always at freedom to
choose this special $SU(2)$ frame) into the last relation in the chain
\p{genvac}, one observes that $<W>$ depends only on
$\theta^\alpha_2$. As a result
\be
\hat{Q}^1_\alpha <W_{\s V}> = 0, \quad \hat{Q}^2_\alpha <W_{\s V}>
\neq 0~,
\ee
i.e. the generator $\hat{Q}^1_\alpha$ annihilates the vacuum, while
$\hat{Q}^2_\alpha$ does not. Hence, $N=1$ supersymmetry
generated by $\hat{Q}^1_\alpha$ is unbroken, and we indeed deal with the
partial breaking $N=2 \rightarrow N=1$ in this case, both
off and on shell.

In this connection, let us recall that the partial breaking nature of the
off-shell $N=2$ supersymmetry Goldstone-type transformation of $W_{\s V}$
\p{modtraV} can be revealed (prior to any on-shell analysis) by shifting
the real auxiliary field $X^{ik}$ of $W_{\s V}$ by a constant triplet
$c^{ik}$  such that the matrix $c^{ik} - (i/2)\; M^{ik}$ is
 degenerate
(in a full analogy with the discussion after eq. \p{shiftWL}). It will be
natural to choose $c^{ik} - (i/2)\; M^{ik} = \hat{x}^{ik}$, so as to have
a one-to-one correspondence between the off- and on-shell partial breaking
regimes. In terms of superfields this amounts to the special choice
of $V^{ik} = V^{ik}_s$ and $W_{\s V_s}\equiv W_s$, such that the vacuum
value of $X_s^{ik}$ vanishes:
\bea
W &=& W_s + \hat{x}^{ik}(\theta_i\theta_k) ~,  \label{special} \\
\delta_\epsilon W_{s} &=& - 2(\epsilon_k\theta_l)\hat{x}^{kl} +
 i(\epsilon Q + \bar{\epsilon}\bar{Q})W_{s}\;.  \label{spectransf}
\eea

Finally, we note that the transformation properties of the $(E,M)$-model
equations of motion under the duality
group $SL(2,{\bf Z})$ become transparent in terms of the dual pair of
superfields $W$ and $\hat{W}_{\s L}={\cal F}_{\s W}$
\bea
&D^{kl}\hat{W}_{\s L}  - \bar D^{kl}\hat{\overline{W}}_{\s L}= 4iE^{kl}~,
& \label{dupair1}  \\
&D^{kl}W  - \bar D^{kl}\overline{W} = 4i M^{kl}~.&   \label{dupair2}
\eea
This pair of equations is covariant under the duality group, provided
that the latter properly rotates $E^{kl}$, $M^{kl}$ through each other.

\setcounter{equation}{0}
\section{Passing to $N=1$ superfields \label{D}}

\hspace{0.5cm}
The properties of the $(E,M)$-mechanism of the $N=2$ supersymmetry
spontaneous breaking were studied in \cite{APT} basically in the
$N=1$ superfield formalism. For completeness, we will briefly
discuss how the above consideration can be translated to this language,
with focusing on the realizations of $N=2$ supersymmetry.

One can pass to the $N=1$ superfield description, expanding $N=2$
superfields in powers of the coordinate $\theta_{\s 2}$. The
coordinate $\theta_{\s 1}$ is assumed to parametrize $N=1$ superspace,
the generators $Q^{\s1}_\alpha$, $\bar Q_{\dot\alpha\s1} =
\bar Q^{\s2}_{\dot\alpha}$ forming the appropriate $N=1$ subalgebra of
$N=2$ supersymmetry.

We will use for $W$ the natural decomposition \p{special}.
The $N=1$ superfield expansion of $W$
reads
\be
W =\hat{\varphi}(x,\theta_{\s 1}) + i\theta_{\s 2}^\alpha
\hat{W}_\alpha(x,\theta_{\s 1})
+ (\theta_{\s 2})^2 A(x,\theta_{\s 1})~.
\label{D10a}
\ee
Then the $N=1$ superfield form of the relation \p{special} is given by
\bea
& \hat{\varphi}=\varphi_s + \hat{x}^{\s 11}
(\theta_{\s 1})^2~,&\label{phispec}\\
& \hat{W}^\alpha =W^\alpha_s - 2i\hat{x}^{\s 12}
\theta^\alpha_{\s 1}~,
 \label{Wspec} \\
& A=(1/4)(\bar{D}_{\s 1})^2\bar{\varphi}_s+
\hat{x}^{\s 22}~, & \label{Aspec}
\eea
where the $N=1$ chiral superfield $\varphi_s $ and the $N=1$ Maxwell
chiral superfield strength $W^\alpha_s$
with a real auxiliary component are coefficients in the $N=1$ superfield
expansion of $W_s$. The superfield $W^\alpha_s $ satisfies
the standard Bianchi identity that can be solved in terms of
the real $N=1$ prepotential $V_s$
\be
   W^\alpha_s =(\bar{D}_{\s1})^2 D^{\alpha\s 1} V_s~. \label{1prep}
\ee

Integrating over $\theta^\alpha_{\s2}, \bar\theta^{\s2}_{\dot\alpha}$,
in the $N=2$ superfield action of the $(E,M)$-model,
one arrives at the $N=1$ superfield action of ref.\cite{APT}. The
analysis of vacuum solutions of the relevant equations of motion has
been already made in ref. \cite{APT}, so we will limit ourselves to
discussing transformation properties of the involved $N=1$ superfields
under the modified $N=2$ supersymmetry in the electric representation.

The $N=1$ superfield components of $W$ possess the standard homogeneous
supersymmetry transformation law
\be
\delta \hat{\varphi}=-i\epsilon^\alpha_{\s 2}  \hat{W}_\alpha+
i(\epsilon_{\s 1}Q^{\s 1}+\bar{\epsilon}^{\s 1}\bar{Q}_{\s 1})
\hat{\varphi}~,
\quad
\delta \hat{W}_\alpha=2i\epsilon_{\alpha\s 2} A+
2\bar{\epsilon}^{\dot{\beta}\s2}\partial_{\alpha\dot{\beta}}\hat{\varphi}
+i(\epsilon_{\s 1}Q^{\s 1}+\bar{\epsilon}^{\s 1}\bar{Q}_{\s 1})
\hat{W}_\alpha~.
\label{stsusy}
\ee
These transformations produce Goldstone-type transformations of
the $N=1$ superfields $\varphi_s$ and $W^\alpha_s$
\bea
&\delta \varphi_s= -2(\hat{x}^{\s11}\epsilon^\alpha_{\s 1} +
\hat{x}^{\s 12}\epsilon^\alpha_{\s 2}) \theta_{\alpha\s 1}
 -i\epsilon^\alpha_{\s 2}  W_{s\;\alpha}
+i(\epsilon_{\s 1}Q^{\s 1}+\bar{\epsilon}^{\s 1}\bar{Q}_{\s 1})\varphi_s
 \label{D12b}&\\
&\delta  W_s^\alpha = 2i(\hat{x}^{\s 22}\epsilon^\alpha_{\s 2}+
\hat{x}^{\s 12}\epsilon^\alpha_{\s 1})
+(i/2)\epsilon_{\s 2}^\alpha [(\bar{D}_{\s 1})^2\bar{\varphi}_s]+
2\bar{\epsilon}^{\dot{\beta}\s2}\partial_{\alpha\dot{\beta}}\varphi_s
+ i(\epsilon_{\s 1}Q^{\s 1}+\bar{\epsilon}^{\s 1}\bar{Q}_{\s 1})
W^\alpha_s~. &
\label{D13b}
\eea

The transformation of the Goldstone scalar superfield
$\varphi_s $ and the $GM$ superfield $W^\alpha_s $ contains
inhomogeneous terms which correspond to the constant translations of
spinor fields. To be convinced in this language that just the
{\it partial} breaking is realized off shell, let us choose the
$SU(2)$ frame \p{frame}. It is easy to observe that in this frame,
with $\hat{x}^{\s22} \neq 0~$, $\hat{x}^{\s12} =
\hat{x}^{\s11} = 0~,$ only $W^\alpha_s$ undergoes an inhomogeneous
shift with the parameter $\epsilon^{\alpha}_{\s 2}$ while
$N=1$ supersymmetry is realized on $\varphi_s$ and $W^\alpha_s$
linearly and homogeneously. Thus $W^\alpha_s $ is the Goldstone
$N=1$ superfield associated with the partial spontaneous breaking
 $N=2 \rightarrow N=1$. This off-shell modification of $N=2$
supersymmetry in the $N=1$ superfield
formalism of the APT model was previously noticed in \cite{BG}.
Our consideration in the previous Sections shows that this phenomenon
arises already after dualization of one of $FI$ terms (this time, the
magnetic one) and actually does not require, on its own range, the
addition of another type of such a term.

Note that the equally admissible choice $\hat{x}^{\s12} = \hat{x}^{\s22}
 = 0~,\hat{x}^{\s11} \neq 0 $ leads to the vanishing of inhomogeneous
terms in the
supersymmetry transformation of $W^\alpha_s$ and the appearance of such
term $\sim \epsilon^\alpha_{\s 1}$ in the transformation of $\varphi_s $.
This means that with this choice the $\epsilon^\alpha_{\s 1}$
$N=1$ supersymmetry is broken, while the $\epsilon^\alpha_{\s 2}$ one is
not. But this does not mean that $\varphi_s $ can be treated as the
corresponding Goldstone superfield in the spirit of ref. \cite{BG2}.
In order to reveal which kind of $N=1$ Goldstone superfield
is actually relevant to one or another pattern of the partial breaking
$N=2 \rightarrow $$N=1$ in the APT model, one should always decompose
$N=2$ superfields over those $N=1$ superfields associated with
the {\it unbroken} $N=1$ supersymmetry. For instance, in the case of the
just mentioned choice of $\hat{x}^{ik}$ a natural decomposition would be
one with respect to the coordinates $\theta_{\s1}, \bar \theta^{\s1}$,
so that the corresponding $N=1$
superfields live on the $N=1$ superspace $(x, \theta_{\s2},
\bar \theta^{\s2})$.
Once again, it is the appropriate $N=1$ gauge superfield which plays the
role of $N=1$ $GM$ superfield in this case.
The same is true for any other choice of $\hat{x}^{ik}$. Thus, the
model under consideration is a kind of `non-minimal' variant of the
nonlinear realization of the partial breaking  with $N=1$
Maxwell superfield as the Goldstone superfield \cite{BG}. Another
version of such a nonlinear realization, with chiral $N=1$ superfield as
the Goldstone one \cite{BG2}, seems to bear no direct relation to the
present model despite the presence of chiral $N=1$ superfield
$\varphi_s$ in $N=2$ $GM$ supermultiplet
(actually, $\varphi_s$ is massive at the points of moduli space
corresponding to the partial breaking).

A modification of $N=2$ supersymmetry can be studied on
the $N=1$ prepotential $V_s$. For the choice \p{entries}, $N=2$
transformation of $V_s$ is as follows
\be
\delta_\epsilon V_s = m(\bar{\theta}^{\s 1})^2 \theta^\alpha_{\s 1}
\epsilon_{\alpha\s 2}
+(i/2)\theta^\alpha_{\s1} \epsilon_{\alpha\s 2}\bar{\varphi}_s
+\mbox{c.c}
+i(\epsilon_{\s 1}Q^{\s 1}+\bar{\epsilon}^{\s 1}\bar{Q}_{\s 1}) V_s~.
\ee
The Lie bracket of these Goldstone supersymmetry transformations contains
pure gauge terms with $(\theta_{\s 1})^2$ and $(\bar{\theta}^{\s 1})^2$,
similarly to the transformation of the vector field in ref. \cite{Fe}.
We will discuss the modification of $N=2$ supersymmetry algebra in more
detail in the next Section.

It would be interesting to study the relation of this
approach to the formalism of nonlinear realization \cite{Ba,BG,BG2}. One
can expect that both approaches are related by an $N=2$ analog of the
nonlinear transformation constructed in \cite{IK} for the case of $N=1$
supersymmetry.

\setcounter{equation}{0}
\section{Consideration in harmonic superspace \label{C}}

\hspace{0.5cm}Harmonic superspace was introduced in ref. \cite{GIK1}
for off-shell description of the gauge,
supergravity and matter $N=2$ supermultiplets.
In ref. \cite{GO} this approach was applied for analysis of spontaneous
breaking of $N=2$ supersymmetry in a general abelian $N=2$ gauge theory.
We will show that it is also helpful for studying dual formulations
of $N=2$ gauge theory and partial $N=2$ supersymmetry breaking.
We use the $SU(2)/U(1)$ harmonics $u^\pm_i$ and the notation
of refs.\cite{GIK1,GI2} for the harmonic and spinor derivatives
in the harmonic superspace, for instance
\bea
& D^{\s++}=\partial^{\s++}-2i\theta^{\alpha+} \bar{\theta}^{\dot{\beta}+}
\partial^{\s A}_{\alpha\dot{\beta}}
+\theta^{\alpha+}\partial^+_\alpha + \bar{\theta}^{\dot{\beta}+}
\bar{\partial}^+_{\dot{\beta}}~, & \label{C1b} \\
& D^+_\alpha=\partial/\partial \theta^{\alpha-}\equiv \partial^+_\alpha~,
\;\;\bar{D}^+_{\dot{\beta}}=
   \partial/\partial\bar{\theta}^{\dot{\beta}-}\equiv
   \bar\partial^+_{\dot\beta}~, & \label{C1c} \\
& D^-_\alpha=-\partial/\partial \theta^{\alpha+} +
2i\bar{\theta}^{\dot{\beta}-}\partial^{\s A}_{\alpha\dot{\beta}}~, &
 \label{C1d} \\
&\bar{D}^-_{\dot{\beta}}=
   -\partial/\partial\bar{\theta}^{\dot{\beta}+} -2i\theta^{\alpha-}
\partial^{\s A}_{\alpha\dot{\beta}}~. & \label{C1e}
\eea
Here
\bea
u^{+i}u^-_i &=& 1~, \quad \partial^{\s ++} = u^{+i}{\partial \over
\partial u^{-i}}~, \quad
\theta^{\pm\alpha} =
\theta^{i\alpha}u^{\pm}_i~,
\quad
\bar\theta^{\pm\dot\alpha} =
\bar \theta^{i\dot\alpha}u^{\pm}_i~,
\nn \\
x_{\s A}^m &=& x^m -i\theta^k \sigma^m \bar{\theta}^l(u^+_k u^-_l +
u^+_l u^-_k )~,\quad \partial^{\s A}_{\alpha\dot{\beta}}=\partial/
\partial x_{\s A}^{\alpha\dot{\beta}}~. \label{HSdef}
\eea

The harmonic-superspace analogue of the constrained superfield \p{A8}
is given by \cite{GIK1,Z1}
\be \label{harmsol}
W_{\s V} \equiv W(V^{\s++}) =(\bar{D}^+)^2\int du_{\s 1}
\frac{V^{\s ++}(z,u_{\s 1})}{(u^+ u^+_{\s 1})^2}=
\int du (\bar{D}^{-})^2 V^{\s++}(z,u)~,
\ee
where a harmonic Green function \cite{GI2} is used.
The superfield
\be
V^{\s++}(z,u) = V^{\s++}(x^m_{\s A}, \theta^{+\alpha},
\bar{\theta}^{+\dot\alpha}, u^{\pm i}) \equiv V^{\s ++} (\zeta, u)
\ee
is the abelian analytic prepotential,
$$
D^+_\alpha V^{\s++}=\bar{D}^+_{\dot{\beta}} V^{\s++}=0,
$$
a fundamental geometric object of $N=2$ gauge theory. It contains $N=2$
vector multiplet and an infinite tower of pure
gauge harmonic components. The gauge freedom of $V^{\s++}$ is given
by the transformations
\be
V^{\s ++} \rightarrow V^{\s ++} + D^{\s ++} \lambda~,  \label{gaugeanal}
\ee
$\lambda = \lambda(\zeta, u)$ being an arbitrary analytic gauge
function. Both $V^{\s ++}$ and $\lambda $, as well as the analytic
subspace $(\zeta^M, u) \equiv
(x^{\alpha\dot\alpha}_{\s A}, \theta^{+\alpha}, \bar \theta^{+\dot\alpha},
u^{\pm}_i)$ are real with respect to some
generalized conjugation \cite{GIK1} (it becomes the ordinary complex
conjugation when applied to conventional $N=2$ superfields).

All the properties of $W$, eqs. (2.6), (2.7), follow in the
harmonic-superspace approach from the analyticity of $V^{\s++}$. The
harmonic superspace form of \p{A7} is as follows
\be
(D^+)^2W_{\s V} - (\bar{D}^+)^2\overline{W}_{\s V} = 0~.  \label{C2}
\ee
The fact that $W$ does not depend on harmonics is expressed
as the condition
$$
D^{\s ++}W = 0\;.
$$
The prepotential $V^{\s++}$ is related to the Mezincescu
prepotential as
\be
V^{\s++} =(D^+)^2(\bar{D}^+)^2[u^-_iu^-_k V^{ik}+\mbox{pure gauge
terms}]~.
\label{megauge}
\ee

Various representations of the $N=2$ Maxwell action with and without
$FI$ terms discussed in the previous Sections can be re-formulated
in terms of harmonic superfields. One can define two sorts of
the analytic prepotentials, `electric' $V^{\s++}$ and `magnetic'
$L^{\s++}$, the latter naturally arising as an analytic Lagrange
multiplier for the constraint \p{C2} in an analog of the intermediate
`master' action introduced in Sect. 3. Below we present a brief account
of the harmonic-superspace version of the APT model considered in Sect.
4.

The basic, holomorphic part of the `master' action has the previous
 form \p{A10}, with $W_{\s V}$ being replaced by an arbitrary
chiral superfield $W$. To appropriately rewrite the Lagrange multiplier
term $S_{\s L}$ in \p{B2}, \p{B3}
we define the dual magnetic chiral superfield strength $W_{\s L}$ on
the pattern of eq. \p{harmsol}
\be
W_{\s L}=(\bar{D}^+)^2\int du_{\s 1}
\frac{L^{\s++}(z,u_{\s 1})}{(u^+ u^+_{\s 1})^2}=
\int du (\bar{D}^-)^2 L^{\s++} (z,u)~.  \label{C1}
\ee
Then this term in the harmonic formalism
can be written as
\bea
&S_{\s L} = -(i/4) \int d^4x d^4\theta du \; W (\bar{D}^-)^2
 L^{ \s++} + \mbox{c.c.}&\nn\\
& =(i/4) \int d\zeta^{(-4)} du \; L^{\s++}\; [(\bar{D}^+)^2
 \overline{W} -(D^+)^2 W ]~, &
\label{C6}
\eea
where
$$
d\zeta^{(-4)} du = d^4x (D^-)^2(\bar{D}^-)^2 du
$$
is the measure of integration over the analytic subspace $(\zeta^M,
u^{\pm}_i)$.
Note that the expression in the square brackets in the second
line of \p{C6} is analytic in virtue of the property $(D^+)^3 = 0$ and
chirality of $W$. Varying $L^{\s ++}$ as an unconstrained analytic
superfeild immdeiately yields the condition \p{C2} which, together with
the chirality of $W$, imply the representation \p{harmsol}.

The `electric' $FI$ term in the electric representation  \p{A10}
can also be written as an integral over the analytic subspace
\be
S_{\s FI}(V) = \int d\zeta^{(-4)} du\; E^{\s ++}V^{\s ++}~,
\label{FIelanal}
\ee
where $E^{\s ++} = E^{ik}u^+_iu^+_k$.
Its `disguised' form pertinent to the master action and defined
in eq. \p{B8}
becomes just \p{FIelanal} after using \p{harmsol} and
decomposing the explicit $\theta$s in \p{B8} in their $\pm$
harmonic projections.
The equation of motion of the abelian model with such a term
is equivalent to the following analytic equation of motion:
\be
(D^+)^2 {\cal F}_{\s W}(W_{\s V}) - \mbox{c.c.} =
\left[\tau (D^+)^2 W_{\s V}  + \tau^\prime D^{+\alpha}W_{\s V}
D^{+}_\alpha W_{\s V}\right] - \mbox{c.c.} = 4iE^{\s++}~.
\label{aneqm}
\ee
It can be obtained by varying the sum of actions \p{A9} (with $W_{\s V}$
given by \p{harmsol}) and \p{FIelanal} with respect to $V^{\s ++}$ (one
should beforehand rewrite the whole action as an integral over the
analytic subspace).

The constraint \p{C2} is modified if we add to the master action
the harmonic version of the magnetic $FI$-term
\cite{GIK1}
\be
S_{m}=-\int d\zeta^{(-4)} du \; M^{\s ++}L^{\s ++}~, \quad M^{\s ++} =
M^{ik}u^+_iu^+_k~,
\label{C8}
\ee
which is an analog of the term (\ref{B13}) (and goes into it after
making use of the magnetic counterpart of the relation \p{megauge}).
The modified form of the constraint contains the $SU(2)$-breaking
constant term (cf. eq.(\ref{B14}))
\be
(D^+)^2 W - (\bar D^+)^2 \overline{W} = 4i M^{\s++}~.
\label{C9}
\ee

At last, the modified magnetic chiral superfield
strength $\hat{W}_{\s L}$ defined by eq. \p{B11} and dual to $W$ obeys
the following harmonic-superspace form of the constraint \p{B12}
\be
(D^+)^2 \hat{W}_{\s L} - (\bar D^+)^2 \hat{\overline{W}}_{\s L} =
4iE^{\s++} \label{C10}
\ee
(it amounts to eq. \p{aneqm} because of the relation
$\hat{W}_{\s L} = {\cal F}_{\s W}(W)$ that follows from the master action
by varying it with respect to $W$). In the rest of this Section we
consider how the magnetic-$FI$ term induced
modification of $N=2$ supersymmetry in the electric representation
affects the transformation properties of $V^{\s ++}$. This modification
was already discussed in Sect. 4 in terms of ordinary $N=2$ superfields
and in Sect. 5 in terms of $N=1$ superfields.

Like in Sect. 5, it will be convenient to use the decomposition
 \p{special} to single out the superfield strength
$W_{\s V}$ subjected to the standard constraint \p{C2} from the object
$W$ satisfying the $M^{\s ++}$-modified constraint \p{C9}
\be
W = W_s + (\theta_k\theta_l)\hat{x}^{kl}~. \label{specharm}
\ee
The first term is expressed according to eq.(\ref{harmsol}) through
the real analytic prepotential $V^{\s++}_s$ with a zero vacuum
expectation value, while the second term contains the constant vacuum
auxiliary field $\hat{x}^{kl}$ defined by eqs. \p{Ysolut} - \p{taurestr}.
 By the modified $N=2$ supersymmetry
transformation \p{spectransf} of $W_s$ we can restore, modulo
the analytic gauge transformations \p{gaugeanal}, the modified
transformation of $V^{\s ++}_s$. As is expected, it is of the
Goldstone type
\be
\delta_\epsilon V^{\s++}_s =-\hat{x}^{kl}\epsilon_k^\alpha
\theta_{\alpha}^+(\bar{\theta}^+)^2 u^-_l  +  \mbox{c.c.}
 + i(\epsilon Q + \bar{\epsilon}\bar{Q})V^{\s++}_s~.  \label{E1}
\ee
For the choice of $SU(2)$ frame as in \p{entries}, the
inhomogeneous term of this transformation is reduced to
\be
 im\;[\; \epsilon_{\s 2}^\alpha
 \theta_{\alpha}^+(\bar{\theta}^+)^2 u^-_{\s2}
-\bar{\epsilon}^{\dot{\alpha}\s2}\bar{\theta}^+_{\dot{\alpha}}
(\theta^+)^2 u^-_{\s1}\;]~.   \label{Spharm}
\ee
It does not contain the parameters of the first
supersymmetry $\epsilon_{\s 1}$ and $\bar{\epsilon}^{\s 1}$, reflecting
the fact that the latter is unbroken under this choice.

It is straightforward to calculate the Lie bracket of the
modified supersymmetry transformations (\ref{E1})
\be
[\delta_\eta,\delta_\epsilon] V^{\s++}_s =
\delta^{stan}_{(\eta,\epsilon)} V^{\s ++}_s +
\delta^{mod}_{(\eta,\epsilon)}V^{\s ++}_s~.
\ee
Besides ordinary $x$-translations and central charge transformations
$\delta^{stan}_{(\eta,\epsilon)}V^{\s ++}_s$, it contains an extra term
which is a special case of the analytic gauge transformation
\p{gaugeanal}
\bea
\delta^{mod}_{(\eta,\epsilon)}V^{\s ++}_s &=&
im [\epsilon^\alpha_{\s2}\eta_{\alpha\s1}-(\eta\leftrightarrow
\epsilon)] (\bar{\theta}^+)^2 u^+_{\s2}u^-_{\s2}
+2im[\eta^\alpha_{\s2}\bar{\epsilon}^{\dot{\alpha}\s2}-
(\eta\leftrightarrow \epsilon)] \theta^+_\alpha
 \bar{\theta}^+_{\dot{\alpha}} u^+_{\s2}u^-_{\s2} \nn \\
&& +\;
2im[\eta^\alpha_{\s2}\bar{\epsilon}^{\dot{\alpha}\s1}-
(\eta\leftrightarrow \epsilon)]
\theta^+_\alpha \bar{\theta}^+_{\dot{\alpha}} u^+_{\s 1} u^-_{\s 2} +
\mbox{c.c.}~.\label{modbrac}
\eea
Defining the appropriate dimension $1$ gauge generators
\bea
&& G\; V^{\s ++}_s \equiv
im\;(\bar{\theta}^+)^2 u^+_{\s2}u^-_{\s2}~, \quad
G_{\alpha\dot\alpha}\; V^{\s ++}_s \equiv 2im
\;\theta_\alpha^+ \bar{\theta}_{\dot\alpha}^+ (u^+_{\s 2}u^-_{\s 2}
- u^+_{\s 1}u^-_{\s 1})~, \nn \\
&& \tilde{G}_{\alpha\dot\alpha}\; V^{\s ++}_s
\equiv 2im \;\theta_\alpha^+ \bar{\theta}_{\dot\alpha}^+
 u^+_{\s 1}u^-_{\s 2}~,
\label{dim1}
\eea
one can write the modification of $N=2$ superalgebra on $V^{\s ++}_s$
as follows (omitting the standard pieces with the 4-translation
and central charges generators)
\be
\{Q^{\s2}_\alpha,Q^{\s1}_\beta\}^{mod} = \varepsilon_{\alpha\beta} G~,
\quad \{Q^{\s2}_\alpha,\bar{Q}_{\dot{\beta}\s2}\}^{mod}
= G_{\alpha\dot\beta}~, \quad
\{Q^{\s2}_\alpha,\bar{Q}_{\dot{\beta}\s1}\}^{mod}
= \tilde{G}_{\alpha\dot\beta}~. \label{modQQ}
\ee
Having found the explicit form of the modified generators $Q$, $\bar Q$,
one can deduce the full modified $N=2$ superalgebra: commuting the
dimension $1$ generators with $Q$, $\bar Q$ produces gauge generators
of dimension $1/2$, etc. In this way one can single out the whole finite
set of the mutually (anti)commuting gauge generators which together with
$Q$, $\bar Q$ form a closed superalgebra. We do not quote it here
 explicitly, limiting ourselves to a few remarks concerning its
 structure. \\

\noindent (i) The standard Poincar\'e or super-Poincar\'e algebras form a
semi-direct product with the corresponding gauge groups treated,
in the spirit of ref. \cite{IvOg},
as groups with an infinite number of generators of the type \p{dim1}.
In the present case some of these
generators appear on the r.h.s. of the basic anticommutators of $N=2$
superalgebra, indicating that we are dealing with a non-trivial
unification of $N=2$ supersymmetry and $N=2$ gauge group. The
existence of such an extended algebraic structure does not
contradict the famous Coleman-Mandula theorem (or any its supersymmetric
generalization), since the two important
assumptions of this theorem, manifest Lorentz invariance and positive
definiteness of the metric in the space of states, cannot be
simultaneously satisfied for gauge theories. \\

\noindent (ii) The above modification should be distinguished from the
well-known modification of $N=2$ current algebra by a constant
`central charge' in the case of partial spontaneous breaking \cite{HLP}.
The latter modification takes place in the APT model too \cite{Fe,Po},
 and it does not influence transformation properties of
the involved $N=2$ superfields, irrespective of whether they are
gauge-invariant or not. \\

\noindent(iii) Gauge transformations in the r.h.s.
of the anticommutator of spinor charges appear also in
the standard supersymmetric gauge theories (without $FI$ terms) as
the result of fixing a gauge. In our case such transformations are
present before any gauge-fixing. \\

\noindent(iv)
Let us emphasize that the modification of $N=2$ transformation law
\p{E1} and, correspondingly, the modification of $N=2$ supersymmetry
algebra \p{modQQ} are unavoidable, they cannot be removed by any
redefinition of the $GM$-prepotential $V^{\s ++}_s$.
However, using the freedom of adding some special gauge
transformations to the `minimal' transformation law \p{E1}, one can
change the structure of the anticommutators \p{modQQ} by gaining some
additional dimension $1$ gauge generators in their r.h.s. \\

\noindent(v) In the modified $N=2$ superalgebra the automorphism
$SU(2)$ symmetry is explicitly broken down to some $U(1)$  due to
the presence of the $SU(2)$-breaking parameters $M^{ik}$ (or $E^{ik}$
in the magnetic representation) as structure constants. \\

\noindent(vi) Due to a non-zero vacuum value of $W|_0 = a$ at the
minimum of the scalar potential corresponding to the partial breaking
of $N=2$ supersymmetry in the APT model, modified $N=2$ superalgebra
necessarily contains, in the anticommutator
$\{Q_\alpha^{\s 1}, Q_\beta^{\s 2}\}$ and its conjugate, a central charge
proportional to the
global $U(1)$ generator of the gauge group. It still
commutes with all gauge generators, is vanishing on $V^{\s ++}$ itself,
 but possesses a non-trivial action on any charged matter multiplet. \\

\noindent(vii) Obviously, the above modification of $N=2$ superalgebra,
in analogy with the central charge deformation just mentioned, should
manifests itself on any charged matter hypermultiplet. This entails
problems with constructing invariant minimal couplings of $V^{\s ++}$
to the analytic $q^{+}$ hypermultiplets. The standard gauge invariant
$q^{+}$ Lagrangian
\be
 \bar q^+ (D^{\s ++} +i V^{\s ++}_s) q^+  \label{qV}
\ee
is not invariant under the modified transformations \p{E1}
and we do not know how to modify the $N=2$ transformation properties of
$q^+$ (and/or the coupling \p{qV}) in order to achieve such
an invariance. Difficulties with coupling of the APT model to an extra
charged matter were earlier noticed in \cite{PP} at the component level.
It is an interesting open problem how to couple $q^+$ to $V^{\s ++}_s$,
and a careful analysis of the representations of the modified
$N=2$ superalgebra should be made in order to solve it.

Finally, we would like to point out that a consistent interpretation of
the partial supersymmetry breaking in the APT model along the lines
of the nonlinear realization approach of refs. \cite{BG}, \cite{BG2} will
require constructing such a realization for the modified $N=2$
superalgebra. One can hope that on this way some unsolved questions
raised in ref. \cite{BG} can be answered. It seems interesting
to seek any other realization of such supersymmetry-gauge algebras
and to reveal their possible stringy origin.

\vspace{0.9cm}
\noindent {\large\bf Acknowledgements}

\vspace{0.2cm}
 We are grateful to J. Bagger, E. Buchbinder, A. Galperin, S. Krivonos,
O. Lechtenfeld, D. L\"{u}st, A. Pashnev and M. Vasiliev for useful
discussions. This work was partially supported by Russian Foundation of
Basic Research (grants RFBR-96-02-17634 and RFBR-DFG-96-02-00180), by
INTAS-grants 93-127, 93-633, 94-2317 and the grant of the Dutch NWO
organization. B.Z. is grateful to Uzbek Foundation of Basic Research for
the partial support
( grant N 11/97 ).


\begin{thebibliography}{99}

\bibitem{APT} I. Antoniadis, H. Partouche and T.R. Taylor,
Phys. Lett. B 372 (1996) 83
\bibitem{Fe} S. Ferrara, L. Girardello and M. Porrati, Phys. Lett.
B 376 (1996) 275
\bibitem{Me} L. Mezincescu, ``On the superfield formulation of
$O(2)$-supersymmetry'', Preprint JINR, P2-12572, Dubna, 1979 (In Russian)
\bibitem{GIK1} A. Galperin, E. Ivanov, S. Kalitzin, V. Ogievetsky and
               E. Sokatchev, Class. Quant. Grav. 1 (1984) 469
\bibitem{BG} J. Bagger and A. Galperin, Phys. Rev. D 55 (1997) 1091
\bibitem{Ba} J. Bagger,  Nucl. Phys. Proc. Suppl. 52 A (1997) 362
\bibitem{GSW} R. Grimm, M. Sohnius and J. Wess, Nucl. Phys.
B 133 (1978) 275 ; \\
M.Sohnius, Nucl. Phys. B 136 (1978) 461
\bibitem{GO} A. Galperin and V. Ogievetsky, Nucl. Phys. Proc. Suppl. 15 B
(1990) 93
\bibitem{SW} N. Seiberg and E. Witten, Nucl. Phys. B 426 (1994) 19
\bibitem{He} M. Henningson, Nucl. Phys. B 458 (1996) 445
\bibitem{CWZ} S. Coleman, J. Wess and B. Zumino, Phys. Rev. 177 (1969)
 2239
\bibitem{BG2} J. Bagger and A. Galperin, Phys. Lett. B 336 (1994) 25
\bibitem{IK} E. Ivanov and A. Kapustnikov, J. Phys. A 11 (1978) 2375;
Nucl. Phys. B 333 (1990) 439
\bibitem{GI2}A. Galperin, E. Ivanov,  V. Ogievetsky and
             E. Sokatchev, Class. Quant. Grav. 2 (1985) 601
\bibitem{Z1} B.M. Zupnik, Sov. J. Nucl. Phys. 44 (1986) 512
\bibitem{IvOg} E. Ivanov and V. Ogievetsky, Lett. Math. Phys. 1 (1976)
309
\bibitem{HLP} J. Hughes, J. Liu and J. Polchinski, Phys. Lett. B 180
(1986) 370
\bibitem{Po} M. Porrati, Nucl. Phys. Proc. Suppl. 55 B (1997) 240
\bibitem{PP} H. Partouche and B. Pioline, Nucl. Phys. Proc. Suppl.
56 B (1997) 322
\bibitem{IKP} E.A. Ivanov, S.O. Krivonos and A.I. Pashnev, Class.
Quant. Grav. 8 (1991) 19

\end{thebibliography}
\end{document}